\documentclass[sigconf,nonacm]{acmart}

\AtBeginDocument{%
  }

\usepackage{tikz}
\newcommand*\circled[1]{\tikz[baseline=(char.base)]{
            \node[shape=circle,draw,inner sep=2pt] (char) {#1};}}
\usepackage{xspace}
\usepackage{listings}
\usepackage{comment}
\usepackage{soul}
\usepackage[normalem]{ulem}
\usepackage{ctable}

\definecolor{dkgreen}{rgb}{0,0.6,0}
\definecolor{gray}{rgb}{0.5,0.5,0.5}
\definecolor{mauve}{rgb}{0.58,0,0.82}
\definecolor{LightCyan}{gray}{0.9}
\definecolor{mGreen}{rgb}{0,0.6,0}
\definecolor{mRed}{rgb}{1.,0.0,0.0}
\definecolor{mPurple}{rgb}{0.58,0,0.82}
\definecolor{mBlue}{rgb}{0,0.0,1}
\definecolor{backgroundColour}{rgb}{0.95,0.95,0.92}
\definecolor{mGray}{rgb}{0.5,0.5,0.5}

\newcommand{\projname}{WarpGuard\xspace}

\lstdefinestyle{inline}{basicstyle={\sffamily\small\spaceskip=0.3em},columns=fullflexible}

\lstdefinestyle{CStyle}{
	commentstyle=\color{mGreen},
	keywordstyle=\color{mBlue},
	stringstyle=\color{mPurple},
	basicstyle=\footnotesize,	
    numbers=left,
    numberstyle=\tiny\color{gray},
    stepnumber=1,
    showstringspaces=false,
    tabsize=1,
	frame=tlbr,
	framesep=0.2cm, 
	framerule=0pt,
	breakatwhitespace=false,         
	breaklines=true,                 
	captionpos=b,                    
	keepspaces=true,                 
	showspaces=false,                
	showstringspaces=false,
	showtabs=false,                  	
	aboveskip=0pt,
    belowskip=0pt,
    frame=tb,
    framerule=1pt,
	language=C,
    xleftmargin=.25in,
    literate={0}{{\textcolor{mPurple}{0}}}{1}%
         {1}{{\textcolor{mPurple}{1}}}{1}%
         {2}{{\textcolor{mPurple}{2}}}{1}%
         {3}{{\textcolor{mPurple}{3}}}{1}%
         {4}{{\textcolor{mPurple}{4}}}{1}%
         {5}{{\textcolor{mPurple}{5}}}{1}%
         {6}{{\textcolor{mPurple}{6}}}{1}%
         {7}{{\textcolor{mPurple}{7}}}{1}%
         {8}{{\textcolor{mPurple}{8}}}{1}%
         {9}{{\textcolor{mPurple}{9}}}{1}%
         {.0}{{\textcolor{mPurple}{.0}}}{2}%
         {.1}{{\textcolor{mPurple}{.1}}}{2}%
         {.2}{{\textcolor{mPurple}{.2}}}{2}%
         {.3}{{\textcolor{mPurple}{.3}}}{2}%
         {.4}{{\textcolor{mPurple}{.4}}}{2}%
         {.5}{{\textcolor{mPurple}{.5}}}{2}%
         {.6}{{\textcolor{mPurple}{.6}}}{2}%
         {.7}{{\textcolor{mPurple}{.7}}}{2}%
         {.8}{{\textcolor{mPurple}{.8}}}{2}%
         {.9}{{\textcolor{mPurple}{.9}}}{2}%
}

\begin{document}

\title[\projname]{\projname: Towards Control-Flow Attestation for Heterogeneous CPU-GPU Execution}

\author{Christian Lindenmeier}
\affiliation{%
  \institution{NVIDIA}
  \city{Santa Clara}
  \country{USA}
}

\author{Meni Orenbach}
\affiliation{%
  \institution{NVIDIA}
  \city{Santa Clara}
  \country{USA}
}

\author{Amro Awad}
\affiliation{%
  \institution{University of Oxford}
  \city{Oxford}
  \country{UK}
}

\author{Fabian Schwarz}
\affiliation{%
  \institution{NVIDIA}
  \city{Santa Clara}
  \country{USA}
}

\author{Fritz Alder}
\affiliation{%
  \institution{NVIDIA}
  \city{Santa Clara}
  \country{USA}
}

\author{Ahmad Atamli}
\affiliation{%
  \institution{NVIDIA}
  \city{Santa Clara}
  \country{USA}
}
\affiliation{%
  \institution{University of Southampton}
  \city{Southampton}
  \country{UK}
}

\renewcommand{\shortauthors}{Lindenmeier et al.}

\begin{abstract}
Heterogeneous CPU-GPU workloads are increasingly used in safety-critical embedded systems, yet no existing approach provides joint attestation of their execution.
Prior Control-Flow Attestation (CFA) techniques focus on CPU-side CFA, while GPU attestation is limited to static, load-time verification and does not provide runtime guarantees.
As a result, runtime attacks on GPU kernels and violations of the CPU-GPU interaction contract remain unaddressed.

We present WarpGuard, the first composite CFA framework for heterogeneous CPU-GPU workloads.
WarpGuard verifies execution against a unified control-flow graph (CFG) that captures both CPU and GPU components.
It extends prior CFA techniques in two ways: it enables runtime CFA of GPU kernels by tracing their execution against kernel-specific CFGs, and it monitors kernel launch events and enforces per-call site policies to detect violations at the CPU–GPU boundary.
These extensions address challenges arising from GPU parallelism and cross-device interactions.

We implement WarpGuard using software-based instrumentation, requiring no specialized hardware or binary modifications.
Our evaluation on an NVIDIA Jetson Orin Nano shows that WarpGuard detects GPU-side control-flow and cross-boundary attacks.
Across microbenchmarks, SPECAccel, and eight TensorRT inference workloads, WarpGuard incurs moderate overheads, suggesting practicality for embedded safety-critical settings.

\end{abstract}
\maketitle

\section{Introduction}\label{sec:intro}
Modern compute workloads increasingly rely on heterogeneous processor combinations: a CPU host orchestrates the overall execution while a GPU accelerator performs data-parallel computation.
This pattern spans scientific computing, machine learning inference, robotics, and autonomous systems~\cite{survey_gpu_cpu,islayem2024hardware,garofalo2025reliabletimepredictableheterogeneoussoc}, and is particularly prevalent in embedded and IoT deployments, where integrated CPU-GPU platforms enable real-time Artificial Intelligence (AI) inference at low power budgets.
As these heterogeneous workloads take on safety-critical roles, verifying their end-to-end execution integrity becomes paramount: a compromised or malfunctioning workload must be detected before it causes harm.

For CPU programs, Control-Flow Attestation (CFA) provides exactly this capability.
CFA allows a remote verifier to cryptographically confirm that a prover executed only control-flow (CF) paths permitted by its Control-Flow Graph (CFG), enabling detection of code-reuse attacks such as return-oriented programming~(ROP)~\cite{roemer2012return} and jump-oriented programming~(JOP)~\cite{bletsch2011jump}.
Since the foundational work of C-FLAT~\cite{c-flat:ccs16}, CFA for CPU-only workloads has matured into a rich research area with solutions across embedded microcontrollers~\cite{litehax:iccad18}, real-time systems~\cite{oat:ieeesp20,scarr:raid19}, and general-purpose platforms~\cite{logbased:ccs19}, covering both software instrumentation and hardware-assisted tracing approaches.

The GPU side of heterogeneous workloads has received comparatively little attention.
Existing GPU integrity approaches, such as SAGE~\cite{sage:atc23}, focus on static, load-time verification: they confirm that a GPU kernel binary is unmodified when loaded onto the device, but provide limited guarantees about the runtime behavior of the loaded code.
Recently, however, Guo~et~al.~\cite{gpu_mem_vuln:usenixsec2024} demonstrated that GPU stack buffer overflows can be exploited to conduct GPU-targeted ROP attacks that are entirely undetectable by load-time binary attestation.
This result shows that GPU runtime CF hijacking is a practical threat today that demands runtime monitoring.
Closing this gap requires overcoming two fundamental challenges.

First, GPU kernels execute thousands of threads concurrently under the Single-Instruction Multiple-Thread (SIMT) model, where threads are grouped into warps that advance in lockstep but can diverge at conditional branches.
Naively tracing CF at per-thread granularity would produce trace volumes orders of magnitude larger than CPU tracing, overwhelming device memory and hampering performance.
Thus, existing CPU CFA techniques do not directly transfer to GPUs: the assumptions of sequential, single-threaded execution do not hold for massively parallel warp execution.

Second, attesting the GPU side in isolation is insufficient.
The CPU application controls which GPU kernel binary is dispatched and how many threads are used, both of which directly affect the computation's execution integrity.
For example, an attacker who can influence the CPU-GPU dispatch boundary can substitute a GPU kernel for another benign one or manipulate the launch configuration to alter execution, none of which leaves a trace in the GPU attestation record alone.
Yet, the computation as a whole might be entirely different from what was intended.
Detecting these CPU-GPU attacks requires binding each GPU execution to the specific GPU callsite on the CPU thread that triggered it, which existing CFA frameworks for either side cannot provide independently.

We present \textbf{\projname}, the first CFA framework that addresses both challenges jointly for heterogeneous CPU-GPU workloads.
\projname traces the CF of both the CPU host application and all dispatched GPU kernels, and correlates each GPU execution with the GPU callsite on the CPU that triggered it.
On the GPU side, \projname exploits the SIMT property to trace at warp granularity: since all active threads in a non-diverged warp execute the same instruction, a single trace entry per warp faithfully captures the warp's CF at a fraction of the per-thread cost.
At the CPU-GPU boundary, \projname records each kernel dispatch, binds it to a content-based kernel identity, and enforces per-GPU launchsite policies specifying which GPU kernels are authorized to execute at each dispatch site and under which launch configurations.
\projname's verifier works on a \emph{composite} CFG unifying both CPU execution and per-kernel GPU CFGs.
We prototype \projname on an NVIDIA Jetson Orin Nano~\cite{nvidia_jetson:online} using software-based instrumentation frameworks DynamoRIO~\cite{dynamorio} and NVBit~\cite{nvbit}, thus not requiring source code, i.e., supporting proprietary AI libraries, or specific hardware modifications enabling widespread adoption.

In summary, our contributions are:
\begin{itemize}
    \item A systematic analysis of the attack surface of CPU-GPU workloads, identifying five attack vectors that no existing system jointly addresses~(Sec.~\ref{sec:cf_attacks}).
    \item The design and implementation of \projname, the first composite CPU-GPU CFA framework, including a composite CFG model, a warp-level GPU tracing mechanism, and CPU-GPU binding~(Sec.~\ref{sec:design}).
    \item A performance evaluation on microbenchmarks, the SPECaccel benchmark suite~\cite{specaccel:online}, and AI IoT inference workloads~\cite{jetsonAIbench:online} showing moderate overhead~(Sec.~\ref{sec:evaluation}).
    \item An end-to-end security demonstration by reproducing the GPU CF hijacking attack of Guo~et~al.~\cite{gpu_mem_vuln:usenixsec2024} and showcasing \projname's practical detection~(Sec.~\ref{sec:evaluation}). 
\end{itemize}

\section{Background}\label{sec:background}

\subsection{Control-Flow Attestation for CPUs}\label{sec:bg:cfa}

CFA enables a \emph{verifier} to obtain cryptographic evidence that a program on a \emph{prover} followed a legitimate execution path~\cite{c-flat:ccs16}.
Unlike Control-Flow Integrity (CFI)~\cite{abadi2005control, cfi_survey}, which enforces a policy at runtime to \emph{prevent} CF deviations locally, CFA passively \emph{records and reports} the execution path to a remote party; enforcement decisions (e.g., alerting or halting the system) are made by the verifier upon receiving the attestation report.
This separation of detection from verification makes CFA suitable for settings where a trusted verifier monitors an untrusted workload by either running locally on the same device in a protected subsystem or entirely remotely.

Most CFA schemes operate in two phases~\cite{c-flat:ccs16,scarr:raid19}.
In the \emph{offline phase}, the verifier pre-computes a reference model of the execution, e.g., the static CFG of each attestable binary.
The CFG is a directed graph whose nodes are basic blocks (BBs) and whose edges represent valid control transfers.
Each binary is identified by a cryptographic hash, which typically includes measurements over the code segment.
In the \emph{online phase}, the prover collects execution traces capturing CF transitions of the program and forwards them, via an authenticated secure channel, to the verifier, which checks conformance either by \emph{CFG replay} (walking the CFG and flagging invalid edges) or \emph{hash-based comparison} against a known-good run.

Trace collection is the primary source of performance overhead.
\emph{Hardware-assisted tracing} uses dedicated CPU features such as Intel Processor Trace or ARM CoreSight to record branches transparently with minimal overhead~\cite{logbased:ccs19, griffin:asplos}.
However, similar features are not publicly available for the GPU side.
\emph{Software-based tracing} instruments the application, either statically at compile time or at runtime via dynamic binary instrumentation (DBI) frameworks such as DynamoRIO~\cite{dynamorio}.
Software tracing is architecture-portable and requires no hardware support, but incurs higher overhead due to the instrumentation hooks executed at each observed instruction.

\subsection{CPU-GPU Heterogeneous Workloads}\label{sec:bg:hetero}
A CPU-GPU heterogeneous workload consists of two components: a \emph{host application} running on the CPU and one or more \emph{kernels} executing on the GPU.
The host manages the full lifecycle: allocating GPU memory, transferring input data, dispatching GPU kernels, and reading back results.
GPU kernels are typically written in C/C++, thus they are susceptible to memory corruptions enabling CF hijacking attacks inside the GPU~\cite{gpu_mem_vuln:usenixsec2024}.
Kernels are compiled to an intermediate virtual ISA (PTX) and translated to the hardware-native binary format (SASS), either ahead of time or via JIT compilation at launch; the resulting images are bundled in a \emph{fatbin} container embedded in the host executable.

At dispatch time, the GPU runtime selects the appropriate SASS image from the fatbin, transfers it to GPU memory, and submits a launch command to the GPU driver.
Crucially, the GPU runtime performs no verification that the dispatched kernel image matches the intended kernel.
Equally, standard CPU CFA traces CF transitions between CPU code locations and has no mechanism to observe what code is loaded into GPU memory or how it executes, making CPU-only CFA fundamentally incapable of attesting CPU-GPU workloads.

\subsection{NVIDIA GPU Architecture}\label{sec:bg:gpu}
\paragraph{Execution Model}
An NVIDIA GPU consists of an array of \emph{Streaming Multiprocessors} (SMs), each capable of executing multiple thread groups concurrently.
The fundamental scheduling unit is the \emph{warp}: a group of 32 threads that execute in lockstep under the Single-Instruction, Multiple-Thread (SIMT) model.
All 32 threads in a warp issue the same instruction each cycle, but each thread operates on its own registers and can follow an independent execution path.
When a conditional branch causes threads within a warp to disagree on the taken direction, the warp \emph{diverges}: the hardware serializes the two diverged groups, executing each in turn while masking inactive threads via a per-warp \emph{active mask} that records which threads are currently executing.

\paragraph{Control-Flow Instructions}
The SASS ISA exposes a rich set of CF instructions relevant to CFA.
{\tt BRA} performs a direct conditional or unconditional branch to an immediate offset; {\tt BRX} performs an indirect branch via a register-indexed jump table.
{\tt CAL} and {\tt JCAL} issue relative and absolute function calls, respectively.
{\tt RET} returns from a called function using an address stored in the thread's local stack frame.
{\tt EXIT} terminates kernel execution.
The divergence management instructions {\tt SSY}, {\tt PBK}, and {\tt PCNT} establish future re-convergence points before a branch, break, or continue; {\tt SYNC}, {\tt BRK}, and {\tt CONT} transfer control to the corresponding re-convergence points.

\paragraph{Memory Model and Local Stack}
Each thread has a private \emph{local memory} region allocated in off-chip global memory, used to store spilled registers and the thread's function call stack.
Stack frames, including return addresses for {\tt RET} instructions, are stored in this per-thread local memory at predictable, fixed offsets within the frame.
Guo~et~al.~\cite{gpu_mem_vuln:usenixsec2024} demonstrated that these offsets are deterministic and accessible to a vulnerable kernel, making it possible for a buffer overflow in one stack variable to overwrite the return address of the same frame, i.e., directly enabling CF hijacking attacks like ROP and JOP inside GPU kernels.
Unlike CPUs, NVIDIA GPU device memory has no execute-permission bit: writable allocations such as those created with \texttt{cudaMalloc} can also be fetched and executed as code, enabling code injection in addition to CF hijacking.
We note that we focus on code-reuse attacks and do not tackle code injection attacks, since they require an orthogonal solution.

\section{Attacks against CPU-GPU Workloads}\label{sec:attacks}
A heterogeneous CPU-GPU workload presents multiple distinct attack surfaces.
In this work, we focus on \emph{control-flow hijacking attacks} that alter the sequence of code executed on the CPU or GPU independently, and on attacks that abuse the CPU-GPU dispatch boundary to manipulate the composite execution without triggering a deviation on either side in isolation.

\subsection{Motivating Scenario}\label{sec:motivating_example}
\begin{figure}[htbp!]
    \centering
    \includegraphics[width=1\columnwidth]{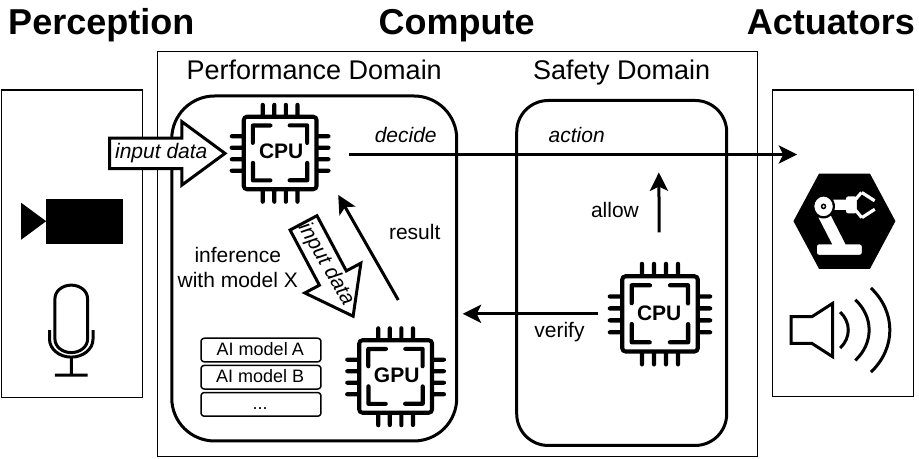}
    \caption{Exemplary AI inference pipeline for a robot: the host CPU selects an AI model based on task context and dispatches it as a GPU kernel in the performance domain; a safety monitor continuously verifies the execution before actuation commands are issued.}
    \label{fig:robot_example}
\end{figure}

We motivate the design of \projname with a scenario from autonomous robotics, a domain where GPU-accelerated AI inference is increasingly deployed on embedded platforms such as the NVIDIA Jetson Orin series~\cite{nvidia_jetson:online}.
Figure~\ref{fig:robot_example} depicts a perception-inference-actuation pipeline: a robot's sensors feed data to a CPU host application, which selects an appropriate AI model, for example, an obstacle detection network or a navigation aid classifier, and dispatches it as a GPU kernel.
The inference result drives a motor command or navigation decision, and different models may be selected at runtime based on operational context (e.g., indoor vs.\ outdoor environment, current mission phase).
A safety monitor, isolated from the performance domain, continuously verifies that the workload executes as intended; if an anomaly is detected, it can halt or slow the robot before an unsafe action reaches the actuators.

In this paper, we focus on software attacks at the application layer: the hardware, OS, GPU device drivers, and CUDA runtime stack are part of the trusted computing base (TCB)~(Sec.~\ref{sec:threat}).
An attacker who gains a foothold at this layer may (1)~exploit memory-safety vulnerabilities in the CPU application or GPU kernels to hijack CF, or (2)~abuse the CPU-GPU dispatch mechanism to alter what code executes on the GPU without triggering any per-side CFG violation.
We now systematically analyze this attack surface and explain why existing attestation approaches are insufficient to detect it.

\subsection{Control-Flow Hijacking Attacks}\label{sec:cf_attacks}

An attacker may exploit a vulnerability in either the CPU or GPU component of the heterogeneous workload to hijack its CF independently.

\paragraph{Class A: CPU Control-Flow Hijacking}
The CPU host application governs the entire workload lifecycle: it acquires sensor data, selects the AI model, launches GPU kernels, and issues actuation commands based on their output.
An attacker who exploits a memory-safety vulnerability in the host application, for example, a stack buffer overflow, can overwrite CF data and construct ROP or JOP chains~\cite{roemer2012return,bletsch2011jump} that diverge from the intended execution path.
This is the standard attacker model for CPU CFA systems; existing frameworks such as C-FLAT~\cite{c-flat:ccs16}, ScaRR~\cite{scarr:raid19}, and ReCFA~\cite{recfa:acsac21} are designed to detect precisely this class of attack.
We design \projname to inherit their guarantees for the CPU side.

\paragraph{Class B: GPU Control-Flow Hijacking}
GPU kernels process attacker-influenced sensor data and are equally susceptible to memory-safety vulnerabilities.
As described in Section~\ref{sec:bg:gpu}, each thread's local memory stores stack frames at predictable, fixed offsets, including the return address for each \texttt{CAL} invocation.
\citeauthor{gpu_mem_vuln:usenixsec2024}~\cite{gpu_mem_vuln:usenixsec2024} demonstrated that this layout can be exploited: a buffer overflow in a kernel's local array can overwrite a saved return address and redirect execution to an attacker-chosen location.

\begin{lstlisting}[language=C, numbers=left, numbersep=4pt, commentstyle=\color{mGreen},
  morekeywords={scalar\_t, __device__},
  caption={Buffer overflow in a matrix-vector multiplication kernel. When the
    attacker controls \texttt{n}, the derived tile count \texttt{ncols} may
    exceed the local buffer capacity, corrupting a saved return address.
    },
  label={lst:gpu-overflow}]
__device__ void matvecmul(scalar_t *T, scalar_t *V,
                         scalar_t *R, int m, int n) {
  scalar_t arr_local[64];
  int ncols = n / blockDim.x;
  int col0  = blockIdx.x * blockDim.x + threadIdx.x;
  for (int k = 0; k < ncols; k++) {
    /* no bounds check: overflow possible */
    arr_local[k] = R[col0 * ncols + k];
  }
  /* RET with corrupted return address */
}
\end{lstlisting}

Listing~\ref{lst:gpu-overflow} illustrates a concrete instance: a matrix-vector kernel allocates a fixed-size local array (line~3) and fills it using a loop bound derived from the attacker-controlled parameter \texttt{n}.
Because \texttt{ncols} (line~4) is never validated against the buffer capacity, a sufficiently large \texttt{n} causes the loop to write past \texttt{arr\_local} into adjacent stack memory, overwriting the saved return address placed there by the \texttt{CAL} that invoked \texttt{matvecmul}.
By supplying crafted values in \texttt{R} at the overflow positions, the attacker redirects execution to an arbitrary location after the function executes its \texttt{RET}.
No existing CPU CFA system can detect this attack, and prior GPU attestation work such as SAGE~\cite{sage:atc23} only verifies static binary integrity at load time: it checks that the kernel binary is unmodified before launch, but provides no protection against runtime CF hijacking.

\subsection{Abuse of CPU-GPU Interactions}\label{sec:cpu_gpu_attacks}
The most distinctive attack class for heterogeneous workloads arises at the CPU-to-GPU dispatch boundary.
As described in Section~\ref{sec:bg:hetero}, kernel images reside in CPU-addressable host memory and are treated as data by the CPU's CFG, i.e., no CPU branch targets GPU kernel code.
An attacker can therefore alter what code is loaded into the GPU or how a kernel is configured at launch, without producing any deviation in the CPU execution trace.
We identify three sub-classes of such attacks.

\paragraph{Class C: Kernel Manipulation}
The attacker patches the kernel binary image in host memory (or on disk before load) before it is transferred to the GPU by, e.g., replacing specific code sequences.
A sophisticated variant leaves the kernel's CF structure intact, so that GPU CF tracing would produce a trace fully compliant with the kernel's CFG, rendering the attack invisible even to CPU or GPU CFA systems that purely rely on CF traces.
SAGE~\cite{sage:atc23} addresses this class by computing a cryptographic measurement of the kernel binary inside the GPU before execution, i.e., confirming its integrity.
\projname also covers kernel image integrity, however, at the GPU runtime API dispatch boundary.

\paragraph{Class D: Kernel Substitution}
A heterogeneous workload typically ships with multiple GPU kernel binaries targeting different operational modes (e.g., different AI model variants or precision settings).
Even when measuring the identity of the kernel loaded into the GPU, one cannot determine which of the application's kernels was \emph{intended} at a given GPU callsite executed on the CPU.
This gap enables kernel substitution: the attacker remaps which kernel image is dispatched by replacing the intended kernel~A with a different, unmodified kernel~B that the verifier knows as legitimate.
In the robotics scenario~(Sec.~\ref{sec:motivating_example}), an attacker could substitute the obstacle detection model with a legitimate but task-mismatched model that reports ``no obstacle'' on most inputs.
Detection requires a binding between the GPU callsite on the CPU and the set of kernels authorized to execute there.
\projname provides this binding via a per-GPU-callsite policy and by tracing GPU kernel launches.

\paragraph{Class E: Kernel Launch Configuration Manipulation}
Beyond the kernel binary, each GPU kernel dispatch carries hardware-level launch parameters: the grid dimension (\texttt{gridDim}) and block dimension (\texttt{blockDim}) define the number of thread blocks and threads per block, respectively.
These parameters directly influence execution: as shown in Listing~\ref{lst:gpu-overflow}, \texttt{blockDim.x} determines loop bounds per thread (line~4) and per-thread memory access offsets (line~5), so a manipulated \texttt{blockDim} can silently alter the computation without violating the kernel's CFG.
\projname records the launch configuration at each GPU callsite and enables enforcing a per-GPU-callsite configuration policy, closing this gap.

\subsection{Coverage Gap}\label{sec:coverage_gap}
Most CFA approaches target CPU CFA~\cite{c-flat:ccs16,scarr:raid19,recfa:acsac21}, i.e., cover Class~A.
GPU kernel manipulation~(Class~C) is addressed by recent work, e.g., SAGE~\cite{sage:atc23}.
However, to the best of our knowledge, no existing CFA system covers GPU CF at runtime (Class~B), leaving GPU workloads entirely unprotected at runtime.
Beyond that, even a hypothetical system that naively combines an existing CPU CFA solution with an independent GPU CFA solution would still miss Classes~D and~E: a substituted kernel produces a benign-looking per-side trace, and a manipulated launch configuration can alter execution behavior while remaining CFG-compliant.
Detection of all five attack classes requires a composite attestation framework that jointly traces CPU execution, GPU execution, and the dispatch events binding them.

\section{Threat Model}\label{sec:threat}
Our threat model is influenced by our motivating scenario of a robotics platform~(Sec.~\ref{sec:motivating_example}) and is therefore tailored for deployments in embedded or edge environments.

\paragraph{Trust Assumptions}
The following software components are part of the TCB: the OS and GPU device drivers, the GPU runtime stack (e.g., CUDA), the prover, and the verifier.
The prover is the trusted measurement agent that instruments and observes the heterogeneous workload.
Depending on the implementation, it receives CF traces via local IPC channels (e.g., shared memory), which we assume to be trusted.

\paragraph{Adversary Capabilities}
The heterogeneous workload, i.e., both the CPU host application and all GPU kernels, running at the application layer, is untrusted and may contain vulnerabilities.
The adversary may feed malicious data into the compute platform via hijacking or manipulating perception sensors~(Sec.~\ref{sec:motivating_example}) in order to exploit, e.g., memory-safety vulnerabilities, and write into non-intended memory areas.
This includes attacks that overwrite any CF data saved on the stack to hijack the CF of either the CPU host application or any of the launched GPU kernels with the goal to construct ROP or JOP chains~(Sec.~\ref{sec:cf_attacks}).

Also, the attacker may try to tamper with the integrity of GPU kernels by manipulating them at runtime in memory or on the disk before they are passed to the GPU runtime API.
On top of that, we include CPU-sided attacks that leverage write primitives to tamper with the CPU-GPU interaction by manipulating launch configuration parameters passed to GPU callsites executed on the CPU.
This includes changing the kernel image pointer or overwriting \texttt{gridDim} or \texttt{blockDim} values~(Sec.~\ref{sec:cpu_gpu_attacks}).

\paragraph{Out of Scope}
Our main objective is the extension of CFA to heterogeneous CPU-GPU workloads, thus we have an overlap with the scope limitations of other related CFA systems targeting CPUs.   
\projname employs a static CFG-based CFA both for CPU and GPU and does not claim to detect CF bending attacks that tamper with runtime-dependent conditional branch or jump instructions.
However, we note that for the GPU these are extremely rare due to performance optimizations, and we did not encounter them in our evaluation set. 
Furthermore, we keep data-only attacks on the CPU that do not target the GPU kernel image or launch configuration at a GPU callsite but change the CPU applications' semantic CF, out of scope.
However, since \projname is compatible with other CPU CFA approaches, this limitation mostly depends on the chosen CPU CFA scheme's properties.
We also note that, while to the best of our knowledge not shown to be practical, data-only attacks inside the GPU are out of scope.
Additionally, we exclude side-channel attacks and dynamically generated or self-modifying code.
Lastly, \projname does not address the threat of GPU code injection attacks~\cite{gpu_mem_vuln:usenixsec2024} via write-executable memory, as we see this to be an orthogonal challenge.

\section{Design}\label{sec:design}
In this section, we present the design of \projname as the first PoC framework that enables capturing the new attack vectors in heterogeneous compute environments, as discussed in Section~\ref{sec:attacks}.
We begin by describing the functional requirements of \projname, the different design options and their trade-offs, the high-level architecture, and main components of \projname.
Implementation aspects of our proof-of-concept are covered in~Section~\ref{sec:implementation}.

\subsection{Functional Requirements}\label{sec:design:freq}
As explained in Section~\ref{sec:bg:hetero}, CFA of heterogeneous workloads must capture the CF behavior of the CPU, GPU, and the interactions between them.
Thus, any attestation reports generated by \projname must capture the following details.
First, the run-time CF behavior of the CPU-side execution, which captures the executed code path in an order-preserving fashion~(Class~A~attacks).
For each run, the prover must ensure that the execution traces are freshly generated and correspond to the run being attested.
Second, for the GPU-side execution, we need to verify that the path taken by each thread is compliant with the allowed CF but in a scalable fashion, which is challenging due to the highly parallel nature of GPUs~(Class~B~attacks).
Finally, and perhaps most importantly, verifying that at each GPU callsite executed on the CPU side, the expected kernel image together with a benign launch configuration is executed~(Class~C, D,~and~E~attacks).

\subsection{Design Space and Trade-offs}\label{sec:design-space}
CFA frameworks can vary in the resolution of CF tracking, required software and hardware support, and the employed logging mechanism and location.
We now elaborate on different trade-offs with the peculiarities of GPUs in mind.

\paragraph{Control-flow Resolution}
The amount of information logged by a CFA framework depends on what type of CF transitions are captured (where) and how often such transitions are captured (when).
In its most aggressive form, CFA reports capture each CFG edge transition, and hence capture each BB's execution (when).
An example to accomplish this is to insert CF capturing hooks in the last or first instruction of each BB (where).
Naturally, this may induce a high performance overhead; thus, a more relaxed version might capture only the function-level CF under the assumption that a CF hijacking attack will most likely surface at function boundaries.
Given the large number of threads executing in GPUs, the difference in performance for BB vs. function-level is more pronounced, hence we support and evaluate both options for GPU kernels.

\paragraph{Required Support} 
While hardware-assisted approaches on the CPU, e.g., ARM's CoreSight, often offer very low performance overheads, they pose challenges of flexibility and wider applicability.
Additionally, to the best of our knowledge, on the GPU side, no equivalent publicly available tracing facility exists.
Thus, while hardware-assisted CFA tracing for CPUs might be an option, the unavailability on the GPU side makes it ill-suited for heterogeneous compute environments, and we opt for a purely software-based instrumentation in our PoC~(Sec.~\ref{sec:implementation}).
Additionally, to support proprietary closed-source libraries (e.g., TensorRT), we opt for a dynamic binary instrumentation tool given its flexibility and potential wider applicability.

\paragraph{Tracing Responsibility}
Given the serial nature of CPU tracing, the tracing responsibility is often held by the executing thread itself, i.e., one trace event per CF event per thread.
However, for GPU kernels, there are usually thousands of GPU threads, and therefore, it is intractable to create and process this massive number of traces corresponding to each GPU thread.
Fortunately, our key insight is that due to the nature of the SIMT execution model~(Sec.~\ref{sec:bg:gpu}), \emph{active} threads within each executing warp follow the exact same CF.
As a result, for the GPU side, we opt for dynamically selecting a \emph{lead thread} which is responsible for tracing the whole warp.
This \emph{warp-level tracing} allows us to significantly reduce the amount of CF traces, making CF tracing on the GPU feasible from a performance perspective without sacrificing security guarantees.

\paragraph{Logging Location}
Execution traces from the CPU and GPU must be logged both efficiently and securely, but the two requirements pull in opposing directions depending on where the buffer resides.
On the CPU side, secure logging buffers can be realized within a trusted execution environment.
No equivalent mechanism is readily available on the GPU.
However, simply writing each GPU trace record directly to host memory at the moment of generation would serialize every warp's trace operation across the CPU-GPU interface, inducing extreme overheads.
We therefore adopt a local tracing strategy as a trade-off: CPU traces are generated in host memory, while GPU traces are generated in GPU-local memory but periodically flushed to the host in batches with the frequency being a user-controlled knob.
We acknowledge that this is a probabilistic rather than cryptographic guarantee, and discuss hardware extensions that could close this gap fully in Section~\ref{sec:discussion:hardware_proposal}.

\subsection{High-level Architecture}\label{sec:design:arch}
In this section, we first describe general assumptions about established CPU CFA on which \projname builds; then, we explain our extensions to integrate the GPU.
Aspects of CPU profiling are detailed in~Section~\ref{sec:cpu_design} while GPU profiling with warp-level tracing is explained in~Section~\ref{sec:gpu_design}.
Sections~\ref{sec:prover} and~\ref{sec:verifier} revolve around \projname's prover and verifier respectively.

\subsubsection{General Assumptions}\label{sec:design:assumptions}
\projname follows the well-established CFA two-party architecture~\cite{c-flat:ccs16}: a \emph{prover} running on the target device and a \emph{verifier} that assesses execution integrity.
However, in contrast to CPU-only CFA, \projname's prover encapsulates the complete execution of a heterogeneous workload that offloads computation to a GPU, collecting CF traces from both the CPU and GPU components and forwarding them securely to the verifier.
The verifier analyzes these reports against a pre-established reference model and raises an alarm on any detected deviation.
In the canonical deployment, the verifier is a remote trusted entity; in our embedded or edge scenario (Sec.~\ref{sec:motivating_example}), it may be co-located on the same device within a protected subsystem.

\paragraph{Authenticity and Freshness}
\projname builds on top of well-established CFA challenge-response protocols established in prior work~\cite{c-flat:ccs16}: the verifier issues a per-session challenge to the prover before any trace data is exchanged, and the prover's reports are bound to that challenge, preventing an adversary from replaying a previously captured attestation session as a fresh execution.
We assume the OS as our root-of-trust (RoT)~(Sec.~\ref{sec:threat}) and rely on it to provide a fresh and authentic communication channel between the prover and the verifier.
This is consistent with the TCB assumptions of CPU-only CFA systems~\cite{c-flat:ccs16, scarr:raid19, recfa:acsac21}, and \projname inherits those guarantees.
We assume that the extensions introduced by \projname run within the same authenticated session, binding GPU-side evidence to the same challenge-response context as the CPU traces.

\paragraph{CPU Binary Integrity}
We further assume that the CPU application binary is attested at load time as part of the challenge-response handshake: the OS-trusted loader measures a cryptographic hash $hash_\text{cpu}$ over the application's code segments and includes it in the initial prover report.
The verifier uses $hash_\text{cpu}$ to select the correct $CFG_\text{cpu}$ from its reference model, ensuring runtime traces are checked against the CFG of the binary actually running on the device.
We note that in CPU CFA schemes, GPU kernel binaries are not checked for integrity, which is a contribution of \projname.

\begin{figure}[htbp!]
    \centering
    \includegraphics[width=0.9\columnwidth]{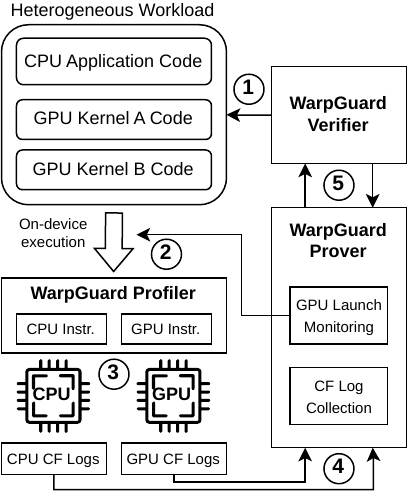}
    \caption{High-level architecture of \projname.}
    \label{fig:design-arch}
\end{figure}

\subsubsection{Extending CFA to GPUs}
We depict the high-level architecture and workflow of \projname in Figure~\ref{fig:design-arch}.
\projname comprises three main components (profiler, prover, and verifier) that operate in five main steps.
Initially, in a measurement step~\circled{\textbf{1}}, which is offline, i.e., before the actual execution on a device, we generate the verifier's reference model in a trusted system using an integrity-verified version of the workload.
\projname processes the heterogeneous workload and divides it into CPU application binaries and GPU kernel binaries.
For the CPU binary, we generate the static $CFG_\text{cpu}$ with its CF edges.
Similarly, we generate the set of static GPU CFGs, i.e., $CFG_\text{gpu,n}$, for each of the $n$ kernels.
We use cryptographic hashes as identifiers for both the CPU and each GPU CFG.
Lastly, we define so-called \emph{GPU launchsite policies}: for each GPU callsite in the CPU binary, a policy lists the authorized GPU kernel hashes and launch configuration values.
In combination, all three artifacts comprise a composite $CFG_\text{comp}$ which our verifier uses as the reference model~(Sec.~\ref{sec:verifier}).
In a second step~\circled{\textbf{2}}, the workload gets executed on a device.
While the launch of the CPU application is handled via established mechanisms~(Sec.~\ref{sec:design:assumptions}), \projname's prover has the capability to intercept GPU kernel launches, which is crucial to validate the integrity of the executed binary but also to record the precise GPU callsite on the CPU.
\projname's prover will also record the kernel launch configuration values and send all this information in a so-called \emph{launch trace} to the verifier.
The launch trace effectively binds a GPU kernel's execution to its GPU callsite~(Sec.~\ref{sec:prover}).
In a third step~\circled{\textbf{3}}, the CPU and GPU parts of the heterogeneous workload get instrumented by \projname's profiler.
Since we opted for a software-based CFA approach, this step entails disassembling both CPU and GPU binaries and inserting logging routines at either BB- or function-level resolution.
Next, both the instrumented CPU and GPU components will be executed on their respective processors and leave traces of CF events in their logging buffers.
In a fourth step~\circled{\textbf{4}}, both CF logs are collected by \projname's prover and sent in batches to the verifier.
In the last fifth step~\circled{\textbf{5}}, \projname's verifier receives the traces and verifies the composite execution of the CPU and GPU under defined GPU launchsite policies using $CFG_\text{comp}$.

\subsection{CPU Profiling and Tracing}\label{sec:cpu_design}
\projname's CPU profiler instruments the CPU application binary at BB granularity, similar to related work~\cite{c-flat:ccs16}.
Before each CF instruction \projname inserts an inline hook that records two values: the instruction offset of the current BB within the binary ($bb\_id$) and the OS-level thread identifier ($thread\_id$).
Together, these two fields allow the verifier to reconstruct the per-thread execution sequence and check it against $CFG_\text{cpu}$.
The complete CPU trace stream $\mathcal{T}_\text{cpu}$ is the sequence of all such flushed messages across all threads.

\paragraph{GPU Callsite Logging}
A key aspect in \projname is the binding of CPU and GPU traces via GPU launch monitoring.
To this end, \projname extends the CPU instrumentation with a dedicated GPU callsite logging mechanism.
\projname scans the CPU binary for calls to GPU APIs (e.g., \texttt{cudaLaunchKernel}) and instruments them with a hook that, at the moment of invocation, logs the call instruction's offset within the binary ($addr\_id$) together with the issuing thread's $thread\_id$ into a separate GPU launchsite buffer.
When the prover later intercepts the GPU kernel dispatch, it reads this buffer to recover the $(thread\_id, addr\_id)$ pair identifying the precise GPU callsite on the CPU.

\subsection{GPU Profiling and Tracing}\label{sec:gpu_design}
While CPU-sided CF tracing is well-established, GPU-sided CF tracing comes, due to the GPU's massively parallel architecture, with unique peculiarities.
We now describe \projname's approach of GPU kernel instrumentation, our compact GPU trace layout, and also the reasoning behind warp-level tracing.

\paragraph{Kernel Instrumentation}
\projname instruments GPU kernel binaries either at BB or function-level resolution: a hook is placed at the first instruction of every BB in the kernel root function and all its related device functions; or only on the first BB for function-level instrumentation, respectively.
The BB-entry strategy reconstructs the executed CF path from a sequence of (prev-BB, curr-BB) pairs: every taken inter-block transfer causes the destination block's entry hook to fire.
As depicted in Table~\ref{tab:gpu-record-fields}, each trace includes an $addr\_id$ field that allows mapping it to a specific node in the kernel CFG.
Divergence-management instructions (e.g., \texttt{BSYNC}) mark future re-convergence targets but do not themselves terminate BBs; they are not instrumented independently, but the BB that contains them is traced like any other.
A re-convergence event, therefore, appears in the trace as a new BB entry under the merged active mask, which \projname's verifier handles through its per-warp divergence-tracking state~(Sec.~\ref{sec:verifier}).

\begin{table}[t]                                                               
  \centering                             
  \caption{\projname's compact GPU trace record. Each record is 16 bytes: two packed 64-bit fields (\texttt{addr\_id} and \texttt{warp\_key}).}                         
  \label{tab:gpu-record-fields}      
  \begin{tabular}{llrp{3cm}}
  \toprule     
  \textbf{Field} & \textbf{Subfield} & \textbf{Bits} & \textbf{Description} \\
  \midrule          
  \texttt{addr\_id} & \texttt{f\_idx}         & 32 & Index of the device function within the kernel's related-function set. \\                       
                      & \texttt{bb\_offset}     & 32 & Byte offset of the BB entry from the start of its function. \\
  \midrule                                                                                                                                                             
  \texttt{warp\_key}  & \texttt{smid}           & 16 & Identifies the physical SM. \\
                      & \texttt{local\_warpid}  & 16 & Warp ID within the SM.  \\
                      & \texttt{active\_mask}   & 32 & Lane execution mask. \\              
  \bottomrule
  \end{tabular}                                                                                                                                                        
  \end{table}

\paragraph{Warp-level Tracing}
Because GPU kernels launch thousands of threads simultaneously, per-thread trace logging is infeasible: in the non-diverged case, all active threads in a warp execute the same instruction, so recording one entry per thread yields 32 identical copies with no additional security benefit~(Sec.~\ref{sec:design-space}).
\projname therefore traces at \emph{warp granularity}: at each BB entry, only the lowest-numbered active lane, determined via the hardware active-mask register, records a trace entry on behalf of the entire warp, reducing trace volume by up to 32$\times$.
Under divergence, SIMT hardware serializes the diverged sub-groups, executing each while masking the others; the first active lane of every serialized sub-group logs independently, so every execution path taken by any thread is represented in the warp-level trace.
As listed in Table~\ref{tab:gpu-record-fields}, each trace uses a \texttt{warp\_key} which allows the verifier to correlate executed traces to their warps and verify their execution.

\subsection{Prover}\label{sec:prover}
At each GPU callsite, \projname's prover intercepts the GPU runtime API to bind the CPU and GPU execution contexts before the kernel is dispatched.          
The prover reads the CPU profiler's per-thread buffer to retrieve the GPU callsite record $(thread\_id, addr\_id)$ logged by the issuing CPU thread (Section~\ref{sec:cpu_design}), then assembles a GPU launch trace $(ctr,\, hash_k,\, config_k,\, (thread\_id, addr\_id))$ and forwards it to the verifier as part of a stream $\mathcal{T}_\text{launch}$.
Here, $ctr$ is a per-kernel launch monotonic counter provided by the OS-based RoT; $hash_k$ is a deterministic content hash over the (offset, opcode) pairs of every instruction in the root kernel and all transitively reachable device functions; and $config_k$ captures the \texttt{gridDim} and \texttt{blockDim} parameters of the launch.

After measuring and recording the kernel launch, the prover reads the GPU-side logging buffer containing the warp-level trace records $(warp\_key, addr\_id)$ produced during execution.
The prover annotates each batch with $ctr$ to correlate it with the corresponding GPU launch trace and forwards the batches to the verifier as a stream $\mathcal{T}_\text{gpu}$.

\subsection{Verifier}\label{sec:verifier}
The verifier receives three distinct trace streams $\mathcal{T}_\text{cpu},\, \mathcal{T}_\text{launch},\, \mathcal{T}_\text{gpu}$ from the prover and combines them with a composite $CFG_\text{comp}$ to produce the final attestation verdict.

\paragraph{Composite CFG}\label{sec:design:composite_cfg}
\begin{figure}
    \centering
    \includegraphics[width=0.5\columnwidth]{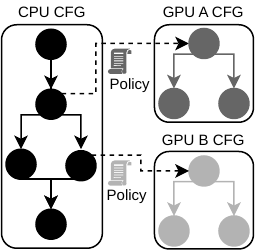}
    \caption{Composite CFG: the CPU CFG (left in black) contains kernel launch nodes that bridge to per-kernel GPU CFGs (right in gray). The GPU launchsite policies bind each GPU callsite on the CPU to a specific GPU CFG.}
    \label{fig:composite_cfg}
\end{figure}
To accurately analyze heterogeneous workloads, a single flat CFG of the CPU application is insufficient~(Section~\ref{sec:attacks}).
\projname uses a \emph{composite} $CFG_\text{comp}$ (Figure~\ref{fig:composite_cfg}) that models the CPU and GPU execution domains separately while linking them at dispatch points via GPU launchsite policies.
We note that the verifier might have multiple CFGs, both for the CPU and GPU side, while cryptographic hashes of the associated binaries are used for identification.
Initially, our verifier uses $hash_{cpu}$ of the CPU binary exchanged during prover communication to identify the CPU-side CFG as the anchor for building the composite CFG.
Thus, we can loosely define $CFG_\text{comp} = (CFG_\text{cpu}, G, P)$ with the set $G=(CFG_\text{gpu,1}, CFG_\text{gpu,2}, CFG_\text{gpu,n})$ and the set $P=(Policy_\text{launch,1}, Policy_\text{launch,2}, Policy_\text{launch,m})$.
$CFG_\text{cpu}$ represents the CF of the initial CPU application in the form: nodes are BBs, edges are statically known branches, jumps, calls, and returns.
Similarly, the set $G$ of $CFG_\text{gpu,n}$ elements represents the CF of GPU kernels, consisting each also of nodes as BBs and edges of statically known CF events.
Note that these are \emph{all} supported GPU kernel CFGs, thus our verifier needs semantic policies to associate them with GPU callsites in $CFG_\text{cpu}$.
The set $P$ holds GPU launchsite policy elements $Policy_\text{launch,x}$ for each GPU callsite $x$ on the CPU; each one holding the following information: (1) $hash_{cpu}$ used to identify the associated $CFG_\text{cpu}$, (2) $addr\_id$ identifying the specific GPU callsite location $x$ within $CFG_\text{cpu}$, (3) $hash_{gpu}$ identifying the expected GPU kernel binary and used to correlate $CFG_\text{gpu,n}$, and (4) $config_k$ defining the expected launch configuration including specification about the expected values for \texttt{gridDim} and \texttt{blockDim}.

\paragraph{Composite Verification}
\projname's verifier uses $CFG_\text{comp}$ in order to verify the holistic CF of a heterogeneous CPU-GPU workload in three steps:
\begin{enumerate}
    \item \textbf{CPU trace verification:} The verifier receives $\mathcal{T}_\text{cpu}$ carrying a sequence of $(thread\_id, bb\_id)$ pairs and verifies compliance with $\mathcal{T}_\text{cpu}$. In detail, for each substream of traces with the same $thread\_id$, this means that (a) every $bb\_id$ corresponds to a valid node in $CFG_\text{cpu}$ and (b) every sequence of $(thread\_id, bb\_id)$ pairs lies on valid edges in $CFG_\text{cpu}$. Non-conformance implies a Class A attack.
    \item \textbf{Launch trace verification:} For every received GPU launch trace $(ctr, hash_k, config_k, (thread\_id, addr\_id)) \in \mathcal{T}_\text{launch}$ the verifier checks compliance with $P$ as: (a)~there exists a policy $Policy_\text{launch,x}$ for callsite $x = addr\_id$, (b)~$hash_k$ is a valid hash in $Policy_\text{launch,x}$, and (c)~$config_k$ is the valid launch configuration in $Policy_\text{launch,x}$. If $hash_k$ is not a valid hash associated with \emph{any} CFG in $G$ it implies a Class C attack. If $hash_k$ is valid in some CFG in $G$ but does not match $hash_{gpu}$ defined in $Policy_\text{launch,x}$, it implies a Class D attack. If $config_k$ from the launch event is not compliant with $config_k$ defined in $Policy_\text{launch,x}$, it implies a Class E attack.
    \item \textbf{GPU trace verification:} For each launch trace verification, the verifier checks the associated GPU traces. We use the value $hash_k$ in combination with the monotonic counter $ctr$ to correlate the associated $\mathcal{T}_\text{gpu}$ and $CFG_\text{gpu}$. For every GPU trace $(warp\_key, addr\_id)$ in $\mathcal{T}_\text{gpu}$ the verifier first checks that $addr\_id$ is a node in $CFG_\text{gpu}$ identified via $hash_k$. Second, for sequences of records with the same $warp\_key$ (which includes the active mask) within one invocation (using $ctr$), consecutive~$addr\_id$~must be connected by a valid CFG edge in $CFG_\text{gpu}$ or be the first instruction of a diverged sub-group (i.e., the traces come from distinct active-mask contexts, indicating a hardware-serialized divergence rather than a sequential transfer). Non-conformance implies a Class B attack.
\end{enumerate}
If all three verification steps hold across all elements observed in $\mathcal{T}_\text{cpu},\, \mathcal{T}_\text{launch},\, \mathcal{T}_\text{gpu}$, the verifier confirms the composite execution as compliant.
Any deviation from these conditions constitutes an attestation failure.

\section{Implementation Details}\label{sec:implementation}
We implement a PoC of \projname on an NVIDIA Jetson Orin Nano (8-core Arm Cortex-A78AE CPU, NVIDIA Ampere GPU, 8\,GB unified memory).
We use dynamic binary instrumentation (DBI) frameworks; DynamoRIO~\cite{dynamorio} for CPU-side instrumentation and NVBit~\cite{nvbit} for GPU-side instrumentation.
We implement \projname's verifier as a standalone application, while \projname's prover is part of our NVBit tool.

\subsection{Instrumentation and Tracing}
\paragraph{DynamoRIO Client}
The CPU side~(Sec.~\ref{sec:cpu_design}) of our PoC is implemented as a DynamoRIO client and uses the readily available API tracing.
CPU trace pairs $(thread\_id, bb\_id)$ accumulate in a per-thread circular buffer; on fill, a flush callback sends the buffered data to the verifier over a TCP socket with a CPU-source header tag.
The size of the buffer is controllable by the user as a trade-off between security and performance.
GPU callsite logging records the module-relative offset of each \texttt{cudaLaunchKernel} callsite into a per-thread shared-memory slot indexed by DynamoRIO thread slot number, so concurrent dispatches from different threads never conflict.
Two mechanisms cover a range of API dispatch conventions.
First, \texttt{\_\_cudaRegisterFunction} is hooked to extract each kernel's runtime stub address into a target table; necessary on AArch64 with static cudart, where all stubs share a single trampoline.
Second, at code-translation time, each BB is scanned, and a clean call is inserted before any direct branch whose target is in the stub table, recording the branch's module-relative offset.

\paragraph{NVBit Tool}
The GPU-side~(Sec.~\ref{sec:gpu_design}) of our PoC is implemented as an NVBit tool and split into a host-side and a device-side part.
The host side intercepts CUDA runtime events via the callback API, disassembles each kernel on first launch, and instruments it at BB or function granularity.       
It iterates over all BBs of the root kernel and every reachable device function, maintaining a per-\texttt{CUfunction} set to skip already instrumented kernels.
A device-side hook is injected at the first instruction of each BB and receives two JIT-baked parameters: a pointer to a per-SM ring buffer and a 64-bit \texttt{addr\_id} encoding the device-function index and the BB's byte offset within that function.
Dividing the ring buffer into per-SM shards turned out to be a key improvement to reduce synchronization overhead.
At each BB entry, the hook calls \texttt{\_\_activemask()} and \texttt{\_\_ffs()} to elect the lowest active lane; all other lanes return immediately, realizing \projname's warp-level tracing.
The elected lane reads \texttt{\%smid} and \texttt{\%warpid} via inline PTX, packs them with the active mask into \texttt{warp\_key}, and writes the record \texttt{(warp\_key, addr\_id)} to its SM's ring buffer.
If the buffer fills, the warp flushes it to the CPU-side prover; when the kernel exits, an \texttt{is\_exit} callback enqueues a flush kernel on the same CUDA stream to drain any remaining records.
The CPU receiver polls all SM doorbells in round-robin order; warp-level trace order is preserved because warp-to-SM assignment is fixed for the duration of a kernel launch.

\subsection{Prover}
Our prover is implemented as part of the NVBit tool and has two responsibilities: launch trace sending and GPU trace forwarding.

On each pre-launch callback, it computes $hash_k$ as an FNV-1a-64 digest over the \texttt{(offset, opcode)} pairs of every instruction in the root kernel and all reachable device functions, cached per \texttt{CUfunction} to avoid re-hashing.
$config_k$ is read from the \texttt{gridDim}/\texttt{blockDim} fields of the \texttt{cuLaunchKernel\_params} parameter struct.
The GPU callsite record is retrieved by matching the issuing thread's Linux TID against the shared-memory region maintained by the DynamoRIO client; the slot is reset to zero to prevent stale data.
The prover sends the assembled launch trace to the verifier over a mutex-protected TCP socket.

A dedicated receiver thread polls the GPU-side per-SM ring buffers; on our evaluation platform, the trace buffer is allocated as CUDA pinned mapped host memory, eliminating \texttt{cudaMemcpy} transfers.
When a doorbell flag signals a completed batch, the receiver forwards it to the verifier, tagged with the monotonic counter.

\subsection{Verifier}
The verifier is a standalone process with dedicated I/O threads that receive CPU, launch, and GPU traces over TCP, and a pool of worker threads that perform CFG replay.
Our implementation focuses on launch and GPU trace verification.
Incoming GPU batches are enqueued to workers using the kernel hash as a sharding key, enabling parallel verification.

\paragraph{GPU CFG Construction}
The GPU CFG is obtained by an offline profiling pass over the target application.
For each kernel, we walk every BB's terminating instruction to derive edges: direct branches (\texttt{BRA}) add one or two edges to statically known targets; blocks ending in indirect transfers (\texttt{BRX}) are marked as wildcard sources, accepting any observed successor at runtime (though we did not encounter any in our evaluation set).
\texttt{CAL} and \texttt{JCAL} encode absolute device addresses and are also treated as wildcard sources; a post-pass then adds a call-to-return edge from each call site to the BB immediately following the call instruction.
Return edges are constructed by collecting the return sites of \texttt{CALL.REL} (per-function) and \texttt{CALL.ABS}/\texttt{JCAL} (global) callers and adding edges from each \texttt{RET} block to the union of the applicable return-site sets.

\paragraph{Launch and GPU Trace Verification}
On receiving a launch trace, the verifier records the received $hash_k$, $config_k$, and GPU callsite address and allocates a fresh per-launch context indexed by $ctr$ to receive subsequent GPU trace batches; the collected launch traces are reported at the end of execution to allow manual inspection against the expected launchsite policies.       

We match the corresponding kernel CFG via its $ctr$ value per GPU trace record.
A node-membership check confirms that \texttt{addr\_id} appears in the CFG node set, and an edge-level check confirms that the transition from the previous block is a valid CFG edge.
While wildcard-source blocks are exempt, we report their occurrence at the end.
When a record's active mask differs from the full warp mask, the verifier opens a per-\texttt{(warp\_id, active\_mask)} divergence slot and tracks that sub-group independently, so a violation in one diverged lane cannot be obscured by another.
Reconvergence is detected when a full-mask record arrives while sub-warp slots remain open; the edge check is suppressed at this transition because the reconvergence target declared by \texttt{BSSY} is not reachable from every diverging path via a single static edge.

\paragraph{Proof-of-Concept Scope.}
For our PoC, CPU traces are collected and sent, but not verified; we defer CPU-side CFG checking to established host CFA implementations~\cite{c-flat:ccs16}.
The GPU launchsite policy verification is not pre-populated.
Instead, the verifier records observed launch traces and prints a summary that requires manual inspection to identify violations.
We defer automating this check to future work.

\section{Evaluation}\label{sec:evaluation}
We test \projname against microbenchmarks, SPECaccel benchmarks, and real-world AI workloads.
Furthermore, we showcase \projname's effectiveness in detecting GPU-based CF attacks.

\subsection{Performance Analysis}\label{sec:eval:perf}
We measure \projname's overhead in various modes designed to isolate individual overhead sources:
\begin{itemize}
    \item $baseline$: Runs the benchmark without any instrumentation.
    \item $DynamoRIO^{bb}_{full}$: Activates only the CPU CFA component tracing on BB-level resolution.
    \item $NVBit^{bb/func}_{empty/full}$: Activates only the GPU CFA component at either BB- or function-level resolution.
    The $empty$ variants insert a no-op NVBit hook to isolate the instrumentation framework's baseline overhead from \projname's tracing logic.
    \item $\projname^{bb/func}_{empty/full}$: Fully activated configuration of \projname.
    The $bb$/$func$ and $empty/full$ variants reference the same options as NVBit's configuration for GPU tracing.
\end{itemize}
For all modes involving NVBit instrumentation, we additionally report the JIT time required to instrument each kernel.

\paragraph{Custom Microbenchmarks}
\begin{figure*}
    \centering
    \includegraphics[width=2\columnwidth]{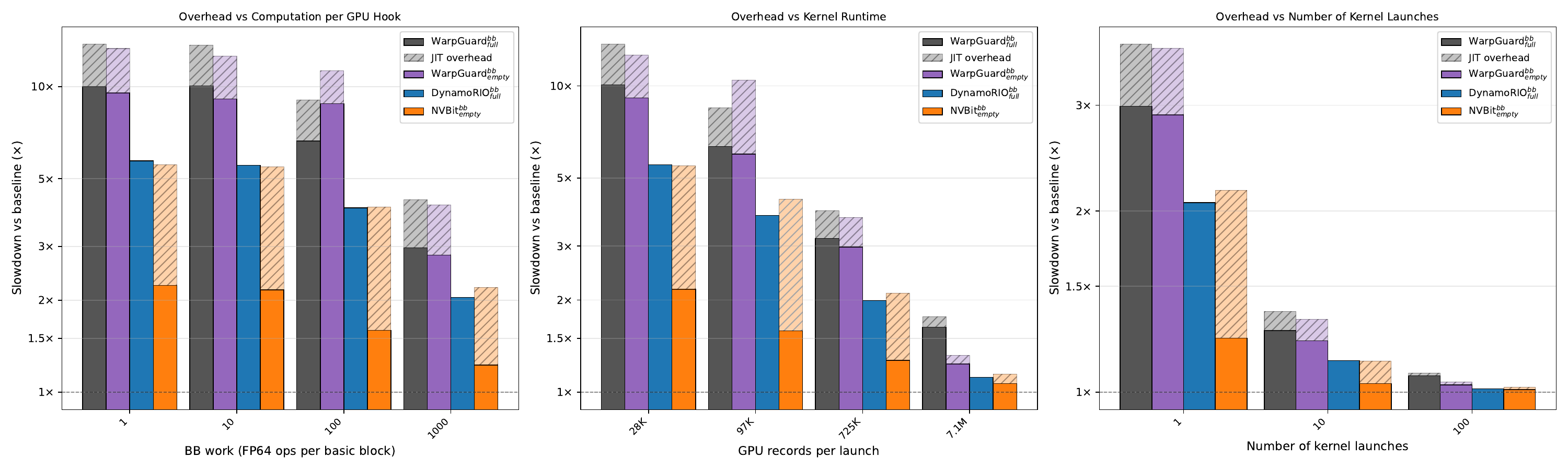}
    \caption{Comparison of different modes of \projname considering our three microbenchmarks. The results show tracing done on BB granularity. Function-level tracing on the GPU shows a similar trend with lower absolute numbers.}
    \label{fig:microbenchmark}
\end{figure*}
In summary, we test our PoC against three custom benchmarks, evaluating different aspects of GPU kernel BB size, GPU kernel runtime, and number of GPU kernel launches.
The results, collected across five runs and averaged, are presented in~Figure~\ref{fig:microbenchmark}.

In the left plot, the microbenchmark kernel is configured with a varying number of FP64 operations per BB and executed ten times; \projname generates ${\approx}281.5K$ GPU records regardless of block size, confirming correct measurement.
With growing computation per BB, $\projname^{bb}_{full}$'s overhead falls since the ratio of work to tracing improves.
In the middle plot, the microbenchmark kernel is configured with a varying number of loop iterations around a fixed-size BB; the increasing GPU record counts on the x-axis confirm correct measurement.
The results show that $\projname^{bb}_{full}$'s overhead falls with longer running kernels as the per-hook cost becomes negligible relative to total kernel work.
In the right plot, the microbenchmark kernel is configured with a fixed BB size and loop iterations; yet, we vary the number of kernel launches.
JIT compilation cost dominates at low launch counts but is amortized as launches increase; $\projname^{bb}_{full}$'s overhead follows the same trend, indicating that \projname scales well with long-running heterogeneous workloads.

\emph{Takeaway:}
Through all microbenchmarks, the gap between \allowbreak $\projname^{bb}_{full}$ and $\projname^{bb}_{empty}$ shows that the dominant overhead stems from the underlying instrumentation frameworks.
$DynamoRIO^{bb}_{full}$ and $NVBit^{bb}_{empty}$ each induce up to ${\approx}5\times$ slowdown independently, and their combination in $\projname^{bb}_{empty}$ accounts for the bulk of the total overhead.
We discuss hardware support that could reduce this framework overhead in Section~\ref{sec:discussion:hardware_proposal}.

\paragraph{SPECaccel Benchmarks}\label{sec:eval:specaccel}
We evaluate \projname on SPECaccel 2023~\cite{specaccel:online}, a suite of GPU-accelerated programs designed for server-class hardware.
To fully stress-test \projname, we deliberately run these benchmarks despite the platform mismatch.
We note that three benchmarks could not execute due to insufficient GPU memory, and we used the \emph{test} workload size.
We run each benchmark five times and average the results.

$\projname^{bb}_{full}$ overhead ranges from $2.4\times$ to $128\times$ (mean $39.3\times$); function-level tracing reduces this to $1.2\times$–$113\times$ (mean $26.7\times$).
Benchmarks with many short-lived, distinct kernels suffer from JIT overhead: spF (7{,}901 launches, $0.56$s base) and swim (122 launches, $0.25$s base) spend $63.7$s and $19.5$s in JIT alone, yielding $128\times$/$113\times$ overhead at $\projname^{bb}_{full}$ and $75\times$/$70\times$ at $\projname^{func}_{full}$.
Long-running kernels with few distinct invocations amortize costs: ilbdc (1{,}000 launches of the same kernel, $17.8$s base) reaches only $2.4\times$ ($\projname^{bb}_{full}$) and $1.4\times$ ($\projname^{func}_{full}$), since its total JIT cost of $0.7$s is negligible against baseline.
For md (3 unique long-running kernels, $2.52$B GPU records), switching to function-level tracing drops overhead from $9.3\times$ to $1.3\times$.
We identified thrashing at extremely high GPU record volumes, indicating that we reach hardware limits.
Consistent with the microbenchmark findings, $\projname^{bb}_{empty}$ closely tracks $\projname^{bb}_{full}$ across all benchmarks, confirming that the instrumentation frameworks account for the dominant share of overhead.

\paragraph{AI IoT Benchmarks}\label{sec:eval:aiiot}
\begin{figure*}
    \centering
    \includegraphics[width=2\columnwidth]{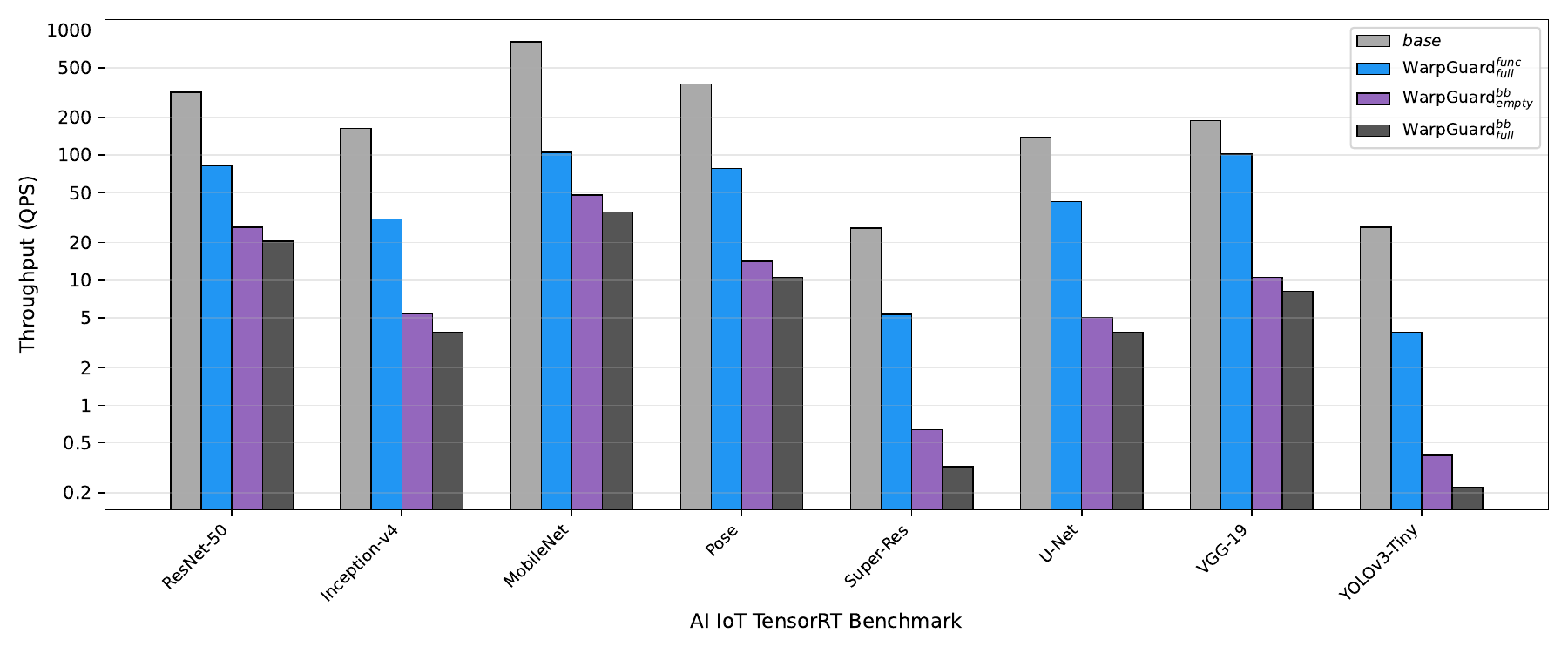}
    \caption{Throughput reached for TensorRT AI workloads by various modes of \projname.}
    \label{fig:AIbenchmark}
\end{figure*}
To evaluate \projname's real-world practicability in the context described in~Section~\ref{sec:motivating_example}, we run eight TensorRT inference engines~\cite{jetsonAIbench:online}, covering common AI IoT tasks: image classification (ResNet-50, Inception-v4, VGG-19), object detection (MobileNet, YOLOv3-Tiny), pose estimation (Pose), image super-resolution (Super-Res), and semantic segmentation (U-Net).
Figure~\ref{fig:AIbenchmark} compares throughput in queries per second (QPS) across modes.
For $\projname^{bb}_{full}$, overhead ranges from 15.5× (ResNet-50) to 120× (YOLOv3-Tiny); switching to $\projname^{func}_{full}$ reduces this to 1.9×–6.9× across all models, retaining throughput levels suitable for edge and IoT deployments: MobileNet retains 104.9 QPS, VGG-19 101.5 QPS, and ResNet-50 82.1 QPS.
Even $\projname^{bb}_{full}$ remains acceptable for ResNet-50 (20.5 QPS) and MobileNet (34.9 QPS) in security-critical settings.
The two outliers, YOLOv3-Tiny ($0.22$ QPS) and Super-Res ($0.32$ QPS), exhibit exceptionally high per-launch GPU record density ($1.2$M and $3.2$M records per launch versus $1.7$K–$42$K for all other models) despite low baseline throughput.
This combination overwhelms two CPU-side components simultaneously: the drain thread cannot forward ring-buffer records fast enough, and the verifier reaches its CPU resource ceiling processing the resulting trace volume.
However, since our PoC prover runs as part of NVBit, i.e., as part of the workload process, it shares the CPU core, thus we believe that shifting this component to a separate multi-core process is an interesting aspect for performance optimizations.

\subsection{Security Evaluation}
Our security evaluation has two parts: practically detecting a GPU CF attack~(i.e., Class~B) and discussing CPU-to-GPU attacks.
Pure CPU-based CF attacks are addressed by related work~\cite{c-flat:ccs16,scarr:raid19,recfa:acsac21,logbased:ccs19}.

\paragraph{Detecting GPU CF Attacks}
We reproduce the PoC attack by Guo~et~al.~\cite{gpu_mem_vuln:usenixsec2024} on our Jetson Orin Nano and adapt it to our scenario.
The attack targets kernel \texttt{k1}, which calls a helper \texttt{sum1} that writes a fixed-size stack array using an attacker-controlled index, i.e., a GPU stack buffer overflow.
A crafted out-of-bounds offset overwrites the saved return address, redirecting execution into \texttt{sum2} (which adds 10 instead of 1 to a buffer and should never be called in a benign execution).
We note that we added a dead call to \texttt{sum2} from \texttt{k1} to force it to be part of the instrumentation based on the CFG.

We run two configurations: (1)~a baseline run without CF hijacking, i.e., below the overflow threshold, and (2)~an attack run triggering the overflow and changing the CF to execute \texttt{sum2}.
In the baseline run, both the verifier's node-membership check and edge-level CFG check pass: all $addr\_id$ values in GPU traces belong to the node set for \texttt{k1} and \texttt{sum1}, and all consecutive warp-level GPU trace pairs $(warp\_key, addr\_id)$ match valid static CFG edges.
In the attack run, the verifier receives $addr\_id$ values corresponding to the \texttt{sum2} function body.
Since the $addr\_id$ values lie inside the statically compiled \texttt{k1} code segment, they correspond to valid nodes in the kernel's device CFG, hence passing the node-membership check.
However, during edge-level CFG check, the verifier immediately identifies a violating edge when execution enters \texttt{sum2} from the \texttt{RET} instruction of \texttt{sum1}.

\paragraph{CPU-to-GPU Attacks}
We reason about \projname's ability to detect Class~C,~D, and E attacks.
In a Class C attack, the adversary modifies the kernel binary image in host memory before GPU API transfer, which will alter the \texttt{(offset, opcode)} pairs over which $hash_k$ is computed.    The resulting $hash_k$ matches no valid entry (assuming no hash collision) in the verifier's offline artifacts.
A sophisticated attacker who modifies code but preserves the CF would still be caught, since $hash_k$ covers all instruction bytes, not just branch targets.
In a Class D attack, the adversary replaces the intended kernel~$A$ with an authorized kernel~$B$ at GPU callsite~$ x$, which leaves both binaries intact. 
However, our verifier looks up $Policy_\text{launch,x}$ and checks that $hash_k$ of $B$ does not equal the authorized kernel hash for that site; since $hash_B \neq hash_A$, again assuming no hash collisions.
In a Class E attack, the adversary tampers with launch configuration parameters (e.g., \texttt{gridDim} or \texttt{blockDim}) to silently change execution without CF deviation.
However, our verifier looks up $Policy_\text{launch,x}$ and checks $config_k$ against the values in $\mathcal{T}_\text{launch}$, thus any mismatch will be detected.

\section{Discussion}\label{sec:discussion}

\paragraph{Trace Encoding}\label{sec:discussion:trace_encoding}
As an alternative to having the prover transmit full execution traces to the verifier, the prover can instead compute and send a cryptographic hash summarizing the trace. 
This hash acts as a compact commitment to the program’s CF, allowing the verifier to confirm execution integrity without processing the entire execution history.

While this approach significantly reduces communication overhead and does not require secure storage for the traces, it introduces certain limitations.
In particular, computing cryptographic hashes over long execution traces incurs unacceptable performance degradation on the prover side, as GPU SMs must spend time computing the hash, which we observed to affect real-time responsiveness. 
Moreover, since the verifier only receives a singular, aggregated hash value, it cannot perform partial or incremental verification, thus sacrificing the ability to localize integrity violations or audit intermediate execution states, crucial for long-running GPU kernels.
Additionally, in practice pre-computing valid hash values for arbitrary inputs can be infeasible.

\paragraph{Future Hardware Recommendation}\label{sec:discussion:hardware_proposal}
A fundamental asymmetry between CPU and GPU security architectures is the absence of hardware-enforced $W\oplus X$, or Data Execution Prevention (DEP), on the GPU side~\cite{gpu_mem_vuln:usenixsec2024}.
Enforcing these would make GPU code injection structurally impossible.
By combining these extensions with \projname's CFA verification, which ensures runtime CF integrity, such a system closes the remaining window for in-GPU CF attacks.

\projname's PoC currently relies on dynamic instrumentation frameworks, notably DynamoRIO~\cite{dynamorio}, NVBit~\cite{nvbit} for both CPU and GPU tracing, respectively.
However, dynamic instrumentation incurs a high performance impact~(Sec.~\ref{sec:eval:perf}).
To mitigate the performance impact on the CPU side, we propose to integrate CFA approaches using Intel PT and ARM CoreSight ETM~\cite{etm:online}, showing overheads below 5\%~\cite{logbased:ccs19, griffin:asplos}.
On the GPU side, no publicly available similar hardware tracing facility exists for NVIDIA GPUs.
We provide two proposals for such future hardware.
First, a cryptographic hash engine that will enable hashing within the GPU directly and communicate the aggregated hash to a verifier.
Alternatively, for a similar trace of visited addresses, future hardware can provide means to trace them directly into CPU memory that would be within the RoT for storage of the prover. 
Regardless on how future GPU, and CPU architectures may expose dedicated tracing hardware, \projname's layered design accommodates such substitutions on GPU and CPU tracing without any major protocol changes.

\paragraph{Applicability to other Architectures and Settings}
\projname's design is not inherently GPU-specific: the composite CFG model, the GPU launchsite policies, and the CPU-GPU binding apply to any system where a CPU host dispatches code to an attached accelerator via a runtime API.
The GPU instrumentation layer would need to be adapted for non-NVIDIA GPUs (e.g., AMD ROCm) or other accelerator types (e.g., TPUs, FPGAs), but the same warp-level tracing principles apply wherever parallel execution under a SIMT model is used.
Our PoC targets the embedded Jetson platform, but \projname's software-based approach is equally applicable to server and HPC settings with dedicated discrete GPUs, as DynamoRIO and NVBit both support those configurations.

\paragraph{JIT-compiled and Dynamic GPU Code}
AI frameworks such as PyTorch and TensorFlow can, in some deployments, use JIT-compiled GPU kernels generated at runtime from Python-level operations.
These kernels are not present in a pre-compiled binary and cannot be covered by a static offline CFG.
JIT compilation makes CFA much harder to deploy correctly, and there are several limitations around precision, security coverage, and performance.
We see JIT support as an important direction for future work.

\paragraph{Trace Buffer Integrity}\label{sec:discussion:tracebuf}
A limitation of software-based GPU tracing is that NVBit's trace buffers are allocated in device global memory, which is accessible to any code running in the same context, including the attacker-controlled kernel being attested.
An adversary who gains control within a GPU kernel could corrupt the trace buffer after a malicious branch executes but before the CPU-sided thread reads it, causing CF events to go unreported.
However, as the buffer can be allocated at arbitrary locations in memory, this imposes a harder barrier for attacking its content due to high address space entropy.
Additionally, the user can define the frequency of trace flushes to the verifier by setting smaller buffer sizes, controlling the attack window and risk.
As proposed above, future hardware extensions could enable secure storage.

\section{Related Work}\label{sec:related}

\paragraph{CPU Control-Flow Attestation}
CFA for CPU-bound workloads has been studied extensively~\cite{c-flat:ccs16,lo-fat:dac17,atrium:iccad17,litehax:iccad18,chase:iccad19,diat:ndss19,attestation:pst19,logbased:ccs19,probability:bigdatase19,radis:sds19,scarr:raid19,cfpa:netsoft19,do-ra:elsevier20,lape:dss20,oat:ieeesp20,arcadis:21,recfa:acsac21,tiny-cfa:date21,cfa_intel_pt:meditcom21,guaranTEE:cloud23,acfa:usenixsec23,ari:usenixsec23,blast:ccs23,isc-flat:rtas23,zekra:asiaccs23,rage:ieeesp24,lightfat:host24}.
C-FLAT~\cite{c-flat:ccs16} pioneered the area with a software-based scheme for embedded systems that uses dynamic binary rewriting to record execution traces and a trusted verifier to replay them against a static CFG.
ScaRR~\cite{scarr:raid19} extended this approach to larger, more complex systems using dynamic binary rewriting with binary-only support.
ReCFA~\cite{recfa:acsac21} introduced resilient CFA using static binary instrumentation.
Hardware-assisted approaches include Log-based CFA~\cite{logbased:ccs19}, which leverages ARM TrustZone and CoreSight to achieve low overhead, and LO-FAT~\cite{lo-fat:dac17}, which uses existing hardware features.
ATRIUM~\cite{atrium:iccad17} targets memory-attack-resilient attestation with dedicated hardware extensions.
Other CPU CFA work addresses constrained IoT devices (Tiny-CFA~\cite{tiny-cfa:date21}, LAPE~\cite{lape:dss20}, LiteHAX~\cite{litehax:iccad18}), collaborative autonomous systems (DIAT~\cite{diat:ndss19}), and operation execution integrity (OAT~\cite{oat:ieeesp20}).
All of these systems operate exclusively on CPU execution and are blind to GPU CF (Class~B) and CPU-GPU attacks (Class~C,~D,~and~E).

\paragraph{GPU Execution Attestation}
SAGE~\cite{sage:atc23} is the closest prior work to \projname on the GPU side.
SAGE uses an SGX enclave to verify that a GPU kernel binary is unmodified before execution, via a pseudo-random memory traversal checksum.
This provides a static binary integrity guarantee but not runtime CF tracing: an unmodified kernel can still have its CF hijacked at runtime via a stack overflow attack~\cite{gpu_mem_vuln:usenixsec2024}.
SAGE also does not address the integration of the GPU attestation with the CPU side.

\paragraph{GPU Control-Flow Attacks}
\citeauthor{gpu_mem_vuln:usenixsec2024}~\cite{gpu_mem_vuln:usenixsec2024} demonstrated that GPU kernels are vulnerable to return-address overwrite attacks due to predictable stack frame layouts in per-thread local memory and the absence of a hardware NX bit on GPU device memory.
Their work motivates the need for GPU-side CFA and directly informs our threat model and PoC design.

\paragraph{Control-Flow Integrity}
\citeauthor{abadi2005control}~\cite{abadi2005control} prevents CF deviations by enforcing a policy locally at runtime, without producing a report for a remote verifier.
Many CFI systems have advanced this concept~\cite{bin-cfi:usenixsec13,ccfir:ieeesp13,llvm_cfi:usenixsec14,kcofi:ieesp14,mcfi:pldi14,rockjit:ccs14}. 
\projname adopts the CFA model with passive recording and remote verification instead. 
This is because CFA is suitable for settings where enforcement decisions are made by an external monitor. 
To our knowledge, no CFI mechanisms exist for GPU execution, leaving GPU kernels without even local CF enforcement before \projname.

\paragraph{GPU Trusted Execution Environments}
An alternative to CFA is hardware-enforced isolation: Graviton~\cite{graviton:osdi18}, HIX~\cite{hix}, and XpuTEE~\cite{xputee} provide GPU TEEs by extending the trust boundary of CPU TEEs (SGX, TrustZone) to the GPU. NVIDIA Confidential Computing embeds a hardware RoT directly on the GPU die~\cite{nvidia_cc}.
These approaches offer strong isolation but require hardware not present on existing embedded GPU platforms, such as the Jetson Orin, and they provide execution isolation rather than an audit trail. 
\projname provides runtime CFA on unmodified existing hardware, complementing TEE approaches where they are available and being the only option where they are not.

\paragraph{Instrumentation Frameworks}
\projname builds on the frameworks DynamoRIO~\cite{dynamorio} and NVBit~\cite{nvbit} for CPU and GPU instrumentation respectively. These are analogous to Valgrind~\cite{valgrind}, Intel PIN~\cite{intel_pin}, ATOM~\cite{atom:pldi94} for CPU and Ocelot~\cite{ocelot} for GPU.

\section{Conclusion}\label{sec:conclusion}
The attestation of heterogeneous CPU-GPU execution is an underexplored area.
Our work presents the first composite CFA framework that jointly attests CPU and GPU execution.
By tracing GPU kernels at warp granularity and binding each kernel dispatch to its originating GPU callsite on the CPU, \projname detects GPU runtime CF hijacking attacks and attacks abusing the CPU-GPU interactions.
Operating entirely with software, our PoC implementation achieves moderate performance overhead, enabling adoption for security-critical scenarios on embedded devices.
Our results demonstrate that composite CPU-GPU CFA is feasible today and we hope \projname serves as a foundation for future hardware-assisted approaches that reduce the remaining instrumentation overhead.

\section*{Acknowledgments}
This work has received funding from the European Union’s Horizon Europe research and innovation programme under Grant Agreement No. \href{https://cordis.europa.eu/project/id/101167904}{101167904} (CASTOR).
Views and opinions expressed are however those of the author(s) only and do not necessarily reflect those of the European Union or the European Commission.
Neither the European Union nor the granting authority can be held responsible for them.

\bibliographystyle{ACM-Reference-Format}
\bibliography{strings, refs}

\end{document}